# Integrated perovskite lasers on silicon nitride waveguide platform by cost-effective high throughput fabrication

Piotr Jacek Cegielski,[1] Stefanie Neutzner,[2] Caroline Porschatis,[1] Holger Lerch,[1] Jens Bolten,[1] Stephan Suckow,[1] Ajay Ram Srimath Kandada,[2] Bartos Chmielak,[1] Annamaria Petrozza,[2] Thorsten Wahlbrink[1] and Anna Lena Giesecke,[1]*

[1]*AMO GmbH, Otto-Blumenthal-Straße 25, 52074 Aachen Germany*
[2]*Center for Nano Science and Technology @Polimi, Istituto Italiano di Tecnologia, via Giovanni Pascoli 70/3, 20133 Milan, Italy*
*giesecke@amo.de

**Abstract:** Metal-halide perovskites are a class of solution processed materials with remarkable optoelectronic properties such as high photoluminescence quantum yields and long carrier lifetimes, which makes them promising for a wide range of efficient photonic devices. In this work, we demonstrate the first successful integration of a perovskite laser onto a silicon nitride photonic chip. High throughput, low cost optical lithography is used followed by indirect structuring of the perovskite waveguide. We embed methylammonium lead tri-iodide ($MAPbI_3$) in a pre-patterned race-track microresonator and couple the emitted light to an integrated photonic waveguide. We clearly observe the build-up of spectrally narrow lasing modes at room temperature upon a pump threshold fluence of 19.6 $\mu Jcm^{-2}$. Our results evidence the possibility of on-chip lasers based on metal-halide perovskites with industry relevance on a commercially available dielectric photonic platform, which is a step forward towards low-cost integrated photonic devices.

## 1. Introduction

Since the two first reports on solid-state hybrid perovskite based solar cells in 2012 [1,2], there has been a world-wide growth of research on these materials, including their implementation in promising light-emitting devices [3,4]. In particular, these novel systems are also being explored as active materials for high quality lasers albeit fabricated via solution processing methods [5]. The first optically pumped laser based on hybrid perovskite semiconductors was made of a polycrystalline thin film of Cl-doped $MAPbI_3$ inside a vertical cavity consisting of a dielectric mirror at the bottom and gold evaporated on top [6]. Since then, several examples of perovskite based lasers were reported including nanowire [7], nanoplate [8], distributed feedback (DFB) [9] and disc [10] lasers with thresholds of 220 $nJcm^{-2}$, 37 $\mu Jcm^{-2}$, 40 $\mu Jcm^{-2}$ and 2.75 $\mu Jcm^{-2}$ respectively. With amplified spontaneous emission (ASE) observed via optical pumping of perovskite thin films even with pulses longer than 100 ns [11], these devices show a great promise for continuous wave lasing. In addition, the efficient and balanced ambipolar carrier transport [3] in $MAPbI_3$ suggests the possibility of building electrically pumped lasers.

Silicon nitride integrated photonic platforms are widely used in applications ranging from telecommunication [12,13] to biosensing [14]. Next to the more mature silicon-on-insulator (SOI) technology they are candidates for (on-chip) optical interconnects, which could replace electrical interconnects offering higher bandwidth and higher energy efficiency [15]. Additionally, the dielectric properties of silicon nitride yield high quality integrated photonic devices useable over a broad spectral range, including visible wavelengths [16]. The drawback of both silicon nitride and SOI platforms is that there are no native light sources available for them. Therefore external sources have to be used, leading to increased packaging and/or assembly costs as well as to significant coupling losses and hence higher required energy per transferred bit [17].

Integration of an efficient light source on chip is highly desired. Extensive work has been undertaken to combine active materials with $Si_3N_4$ waveguide wafers by adding additional materials such as III-V semiconductors [18], which require epitaxy or bonding of III-V materials to a silicon nitride waveguide wafer. In another example colloidal quantum dots were used for an optically pumped laser by embedding them in a $Si_3N_4$ disc raised on a Si pillar [19], which required a rather elaborate fabrication. Both approaches could be, in principle, used for integrated light sources, however complexity of fabrication would limit the industrial relevance due to high costs. Perovskites, on the other hand, are processed at low temperatures from solution by simple and scalable methods [20], which enables them to be easily added to existing photonic integrated circuits (PICs), making them a great alternative for on chip light sources for the silicon nitride waveguide platforms.

Previously demonstrated perovskite microlasers [7,8] were obtained by crystal growth, which is problematic for integration with classic, top down large scale PICs. To benefit from straightforward, low-cost integration on silicon nitride PICs the perovskites must be patterned. This is nontrivial as perovskites are not compatible with chemistry and processing required for optical lithography. Recently top-down patterning of $MAPbBr_3$ perovskite by e-beam lithography with poly(methyl methacrylate) (PMMA) resist has been demonstrated [10]. Although remarkable results have been achieved, this method is limited by low throughput of e-beam systems and risk of cross-contaminating other samples processed in the same tools with elements contained within perovskites (such as Pb, Br and I). Moreover, to this moment, no integration with commercial photonic circuitry was demonstrated.

In this work we present a low-cost, high throughput process flow for fabrication of perovskite devices integrated with silicon nitride photonic chips. It can be combined with existing silicon nitride PIC process flows. Also, it can be used as a photonic test platform for various organo-metal halide perovskites or any other solution processed materials with refractive index high enough to obtain a guided mode (e.g. F8BT [21]). As a proof of concept we present a $MAPbI_3$ ring laser integrated with a $Si_3N_4$ photonic integrated circuit.

## 2. Laser design

We chose a racetrack resonator (Fig. 1(a)) as a perovskite feedback structure rather than Bragg gratings and photonic crystals due to its lower lithography resolution requirement (500 nm vs. sub 250 nm) as well as its higher robustness to fabrication errors. Additionally, using a racetrack resonator enables transfer of light, generated by pumping with a focused laser beam, to a silicon nitride waveguide by evanescent field coupling using a vertical directional coupler (Outlined in Fig. 1(a), cross section in Fig. 1(b)). Our device contains two layers: a silicon nitride photonic circuit layer and a perovskite device layer above. The basic building block - the perovskite waveguide - is fabricated by filling the trench etched in $SiO_2$ cladding layer by spincoating liquid perovskite precursor (details in section 3), which leaves also a thin film on the entire chip surface. To obtain well defined waveguides by this method the trenches have to be deep and narrow, so that the thin film do not affect the guided mode, which should be confined closer to the bottom of the trench. This reduces bending losses caused by guiding of light by thin film on the surface and increases coupling efficiency of the vertical directional coupler. The design aimed at ~1% coupling efficiency to $Si_3N_4$ waveguide in order to output measurable power of the generated light and maintain high quality (Q) factor of the resonator.

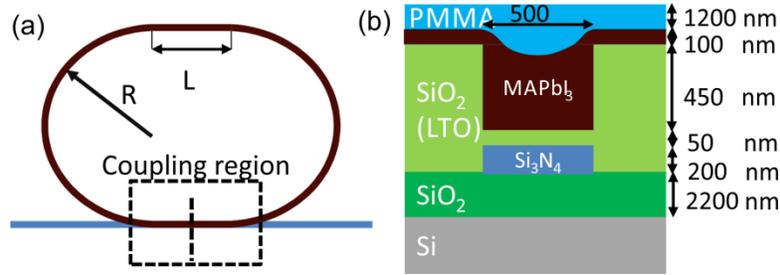

Fig. 1. (a) Sketch of the racetrack perovskite laser. (b) Cross section of the coupling region.

The perovskite waveguide width was set to 500 nm limited by the lithographic tool resolution. The depth of 450 nm was chosen as a tradeoff between confining light away from the surface film and providing enough evanescent field for coupling with the single mode $Si_3N_4$ nitride waveguide (500 nm wide and 200 nm high). Both waveguides support TM and TE fundamental modes. Because perovskite waveguide mode was strongly confined due to high refractive index of $MAPbI_3$ the coupler gap was set to only 50 nm to increase coupling. Additionally, a straight section of length L was added to the ring to enlarge the coupling coefficient. Different types of couplers with varying lengths (2.5 µm, 6.7 µm and 11 µm) were fabricated and tested. A calculation of the group index ($n_g$) of the mode based on refractive index (n) values obtained by P. Loeper and M. Stuckelberger [22] resulted in $n_g \approx 6$. This relatively high value of $n_g$ is due to the strong dispersion of the dielectric function of $MAPbI_3$. Because of that the curvature radius of the racetracks of 15 µm was chosen to ensure that the free spectral range (FSR) between the resonances is high enough to distinguish the resonance peaks (approx. 1 nm) and that the bending loss is low.

### 3. Indirect patterning of perovskite waveguides

For the fabrication of perovskite waveguides an indirect patterning approach is used because $MAPbI_3$ is damaged by the chemicals typically used for optical lithography. In Fig. 2(a-d) the fabrication process of an example perovskite resonator integrated with silicon nitride photonic circuit is schematically shown. Devices were fabricated on 6" silicon wafers with 2.2 µm of thermally grown $SiO_2$. 200 nm of silicon nitride were deposited by a low pressure chemical vapor deposition (LPCVD) process from dichlorosilane and ammonia gases. Silicon nitride waveguides were patterned by optical lithography with a Canon i-line stepper and etched in an inductively coupled plasma reactor using $CHF_3$ and helium gases (Fig. 2(a)). This was followed by a LPCVD deposition of 850 nm of $SiO_2$ (low temperature oxide - LTO) by silane and oxygen reaction. Planarization of the $SiO_2$ surface was done by spin coating spin-on glass (SOG) and subsequent etching in $CHF_3$ plasma. The process cycle of SOG coating and dry etching was repeated twice, leaving only a 20 nm level difference between the LTO surface above and away from silicon nitride waveguides. Hence, about 100 nm of LTO was consumed during these planarization steps (Fig. 2(b)). Next, trenches were patterned by the same lithographic technique and etched in $CHF_3$ and Ar plasma. As no etch stop layer was present, etching was done in steps to carefully control the etch depth with an ellipsometer (Fig. 2(c)). Finally, wafers were diced into separate chips.

For the synthesis of $MAPbI_3$, solvent engineering was used [23]. The precursor solution was prepared by mixing $PbI_2$, methylammonium iodide and dimethyl sulfoxide in dimethylformamide. The solution was spin coated in glovebox conditions on PICs, which were treated with oxygen plasma shortly before. Samples were spun at 4000 rpm and toluene was dripped after 6 s to promote fast crystallization and reduce the average crystal size. Annealing was done in two steps: first at 55 °C for 1 minute and then for 10 minutes at 100 °C. The resulting crystals of $MAPbI_3$ located inside the trench have diameters of approximately 200 nm. In the last step chips were covered with 1.2 µm thick PMMA encapsulation layer. As a result a flipped upside down rib waveguide [24] is obtained(Fig. 2(d)). The SEM cross section of the coupling region of the fabricated devices is presented in Fig. 2(e). The spin-coated perovskite filled the groove up to 320 nm and left an excess layer of 120 nm on top of $SiO_2$. The obtained waveguide supports TE and TM modes (calculated by finite difference method mode solver implemented in OptoDesigner by Phoenix Software), which are almost not interacting with the film on the surface (see insets in Fig. 2(e)).

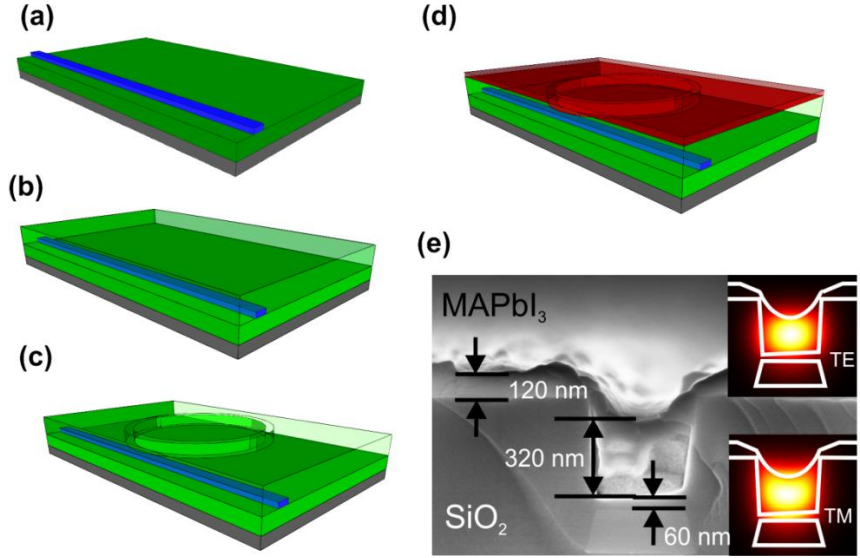

Fig. 2. Sketch of the fabrication process: (a) patterning of silicon nitride waveguide on $SiO_2$, (b) deposition of $SiO_2$ cladding and planarization, (c) etching of trenches in $SiO_2$, d) perovskite deposition via spin-coating. (e) SEM micrograph of the device cross section in the coupling region. Inset: TE and TM mode profile in the perovskite waveguide.

With this method perovskite is applied at the end of the fabrication process and does not undergo any further processing apart from coating with a PMMA encapsulating layer. Therefore it will not be affected by the lithographic process, the cross-contamination risk is eliminated, and also, the type of perovskite used can be changed without need for major adjustments of pattering parameters or etching conditions. In principle, $MAPbI_3$ can be replaced with perovskites emitting light at different wavelengths depending on the desired application or by any other solution processed material with sufficient refractive index contrast with respect to $SiO_2$. Because trenches are etched in the cladding layer, which is the last step of an photonic integrated circuit fabrication our method can be easily combined with existing silicon nitride PIC [25] process flows.

## 4. Experimental setup

The measurement scheme used to collect the emitted light generated either in the racetrack resonator or in the excess perovskite layer on top of $SiO_2$ is shown in the Fig. 3(a). The material was pumped from the top with 120 fs laser pulses at 645 nm focused by a 15 cm lens. The beam was generated by a laser system consisting of a mode-locked Ti:sapphire oscillator (Coherent Micra) in conjunction with a regenerative amplifier (Coherent RegA 900) and an optical parametric amplifier (Coherent 9450). The samples were positioned on a sample chuck mounted on a XY micrometer stage and held by vacuum. To enable alignment of the optical fiber end to the sample edge, a microscope was positioned above the sample, equipped with a 20x objective with a working distance of 1.5 cm. This forced the pump beam to be directed at an angle of ~55° causing a reflection of 10% of the incoming power. We calculated the Gaussian spot size of the elliptically shaped laser spot to be 2477 µm$^2$, using an image of photoluminescence from the excited perovskite. The output was collected by end-fire coupling to a step index multimode fiber (Thorlabs, core 50 µm, 0.22 NA). A multimode fiber was used to relax alignment tolerances, because samples were cleaved by hand after deposition of perovskite resulting in uneven edges. Spectra were measured by a Maya2000 Pro (Ocean optics) spectrometer with 0.4 nm resolution and integration times ranging from 10 to 50 ms. All samples were measured in air at room temperature.

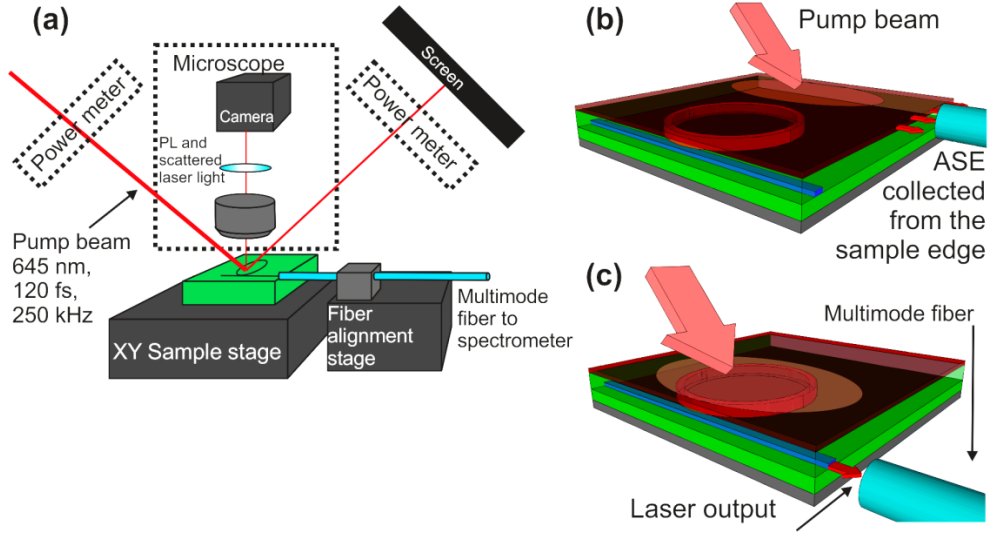

Fig. 3. (a) Measurement setup. (b) Schematic of ASE measurement. The pump beam is moved away from the resonator together with the optical fiber to collect light from the perovskite thin film. (c) Schematic view of end-fire coupling to the laser output. The pump beam (outlined) is focused onto the resonator while the optical fiber is aligned to the silicon nitride waveguide.

## 5. Characterization of the laser

Prior to the characterization of the perovskite resonators we have excited the thin perovskite excess film and collected the generated light from the sample edge (Fig. 3(b)). At low pump power we observed a broad photoluminescence spectrum peaking around 770 nm. After reaching an excitation fluence of 29.4 µJcm$^{-2}$ the typical, spectrally narrowed feature of ASE appeared around 790 nm (Fig. 4(a)). Upon increasing excitation power the ASE peak shifted slightly to longer wavelengths due to band gap renormalization at high excitation densities [26]. When the pump beam was focused on a resonator and the multimode fiber was aligned to collect light from the silicon nitride waveguide (Fig. 3(c)), we could observe the formation of narrow lasing modes upon increasing the excitation power (Fig. 4(b)). Lasing can be recognized [27] by narrowing of the emission spectrum as well as by a clear threshold in the peak emission intensity curve at 19.6 µJcm$^{-2}$ (inset of Fig. 4(b)) for ring resonators with a 6.7 µm long coupling section. At excitation fluences higher than 65 µJcm$^{-2}$ the gain starts to saturate. A resonant mode pattern can be seen in the microscope image of the pumped resonator (Fig. 4(c)). A comparison with the ASE characteristics obtained from the thin perovskite film (inset in Fig. 4(a)) reveals a significant difference in the output intensity curves. The signal from the laser resonator shows a much steeper slope by over two orders of magnitude due to positive feedback of the resonant cavity, which is an additional indication of lasing.

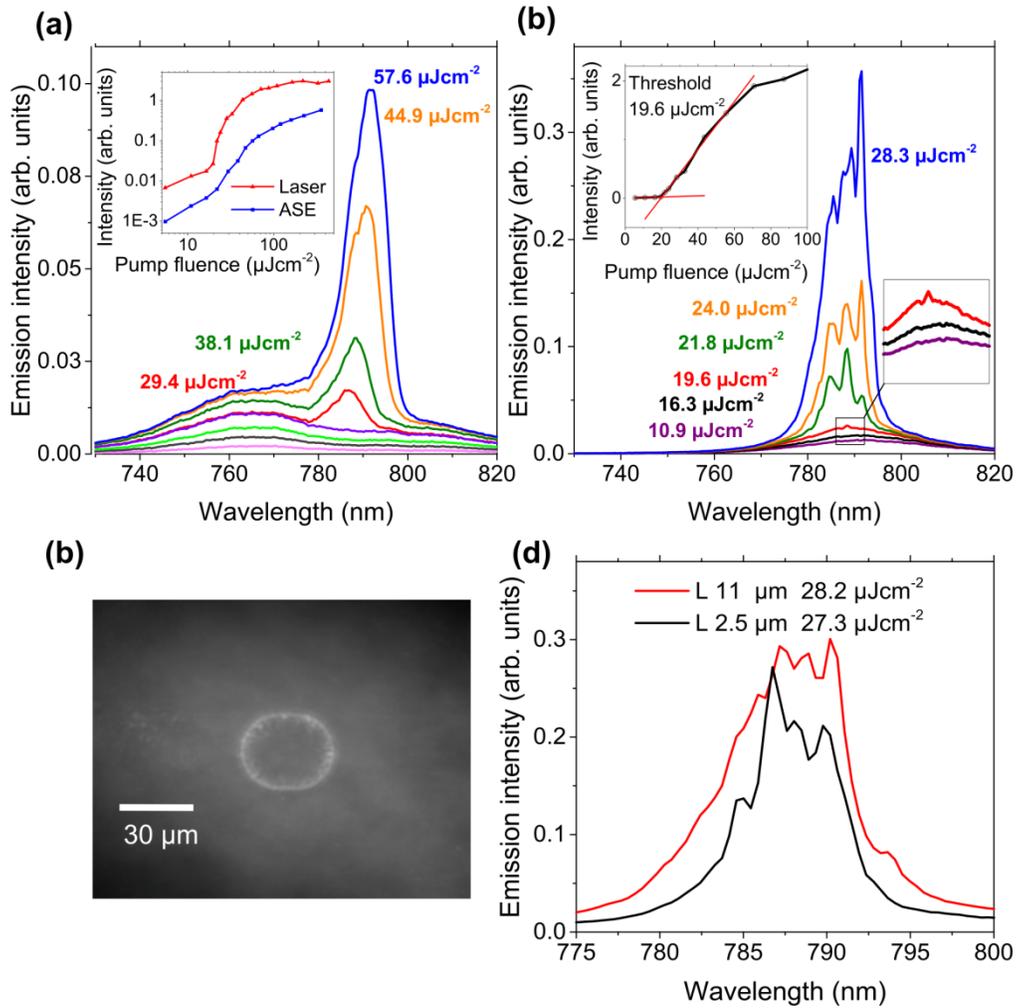

Fig. 4. (a) ASE spectra of the perovskite thin film. Inset: Peak intensity vs. pump fluence. (b) Emission spectra of the integrated perovskite laser (R = 15 µm, L = 6.7 µm) at different excitations Inset: peak intensity vs. excitation fluence with a threshold of 19.6 µJcm$^{-2}$. (c) Image of photoluminescence from excited racetrack resonator. The scattered light of the pump beam was removed with an optical longpass filter. (d) Comparison of emission spectra of resonators with L = 2.5 µm and L= 11µm pumped just above the threshold.

Above 21.8 µJcm$^{-2}$ additional modes of the ring cavity reached their threshold and appeared in the spectrum at 784.6 nm and 791.5 nm (Fig. 4(b)). Upon further increase of excitation, we observed that the mode at 791.5 nm exceeded the intensity of the center mode (788.5 nm), due to red shifting of the gain region of the emission spectrum observed as shifting of ASE described earlier. Additionally, the appearance of a mode at 789.3 nm at the excitation of 28.3 µJcm$^{-2}$ was observed, which can be caused by interaction of the modes due to scattering at the grain boundaries between crystalline domains of the perovskite. Fig. 4(d) depicts the spectral mode distribution of lasers with coupling lengths of L = 2.5 µm and L= 11 µm. The average FSR is 1.73 nm and 1.42 nm respectively, which follows the dependency of the FSR on the resonator length. Devices were measured across several chips with consistent results, indicating that the fabrication technique provides reliable devices. The laser was pumped with 120 fs-pulses, therefore it did not reach steady state in which a single mode would be expected to win over the other modes. Photoluminescence background collected from the Si$_3$N$_4$ waveguide was much weaker than in the ASE measurement at wavelengths shorter than 780 nm due to reabsorption in the perovskite ring because the absorption coefficient of MAPbI$_3$ raises significantly for wavelengths < 780 nm [22]).

The Q factors of the lasing modes just above the threshold, approximated by Lorentzian fitting, were in the range of 600 to 650 while at higher pump fluencies they dropped to values from 100 to 300. This implies high propagation losses in the perovskite waveguide. The microscope image (Fig. 4(c)) confirms these findings, as it shows strong scattering in the resonator caused by grain boundaries and the rough surface of the MAPbI$_3$ waveguide. Based on the approximate resonator Q factor values, assuming low bending loss (0.1 dB/90° bend) and a 1% coupling efficiency of the perovskite to the silicon nitride coupler, the propagation loss of the perovskite waveguide was calculated to be 18.5 dBcm$^{-1}$. The modal net gain coefficient is therefore ~70 cm$^{-1}$, which is in the range reported in [28]: 40-250 cm$^{-1}$, and led to lasing despite very high propagation losses in the perovskite waveguide.

## 6. Summary

We presented the first - to the best of our knowledge - perovskite laser integrated with a silicon nitride photonic integrated circuit. Light generated in the perovskite racetrack microresonator was coupled into a silicon nitride waveguide and measured by end fire coupling to a multimode fiber. The achieved lasing threshold of 19.6 µJcm$^{-2}$ is in the range of the reported values of thin film MAPbI$_3$ lasers [9,29]. We manufactured perovskite racetrack resonators by a fabrication method, which bypasses the incompatibility of perovskites with most of top-down patterning techniques. In addition, the throughput of our method is many times higher than already reported patterning by e-beam [10] or focused ion beam (FIB) [30]. Also, perovskite is deposited last therefore it is not affected by the patterning process. The perovskite composition can be altered without need for any change in the PIC fabrication process flow making it a valuable material testing platform.

Quality factors of the resonators, and thus the laser threshold, can be improved by lowering scattering losses through reduction of the crystal size of the perovskite [31]. The wavelength can be tuned to the telecommunication standard of 850 nm by using tin based perovskites [32]. In conjunction with future developments of electrical injection, our prototypes are one step forward towards a cost-effective, solution processed, industry relevant on chip laser on a dielectric photonic platform.

## Funding

European Union's Horizon 2020 research and innovation programme under the Marie Sklodowska-Curie grant agreement No 643238 (Synchronics) and PPP No 688166 (Plasmofab).